\documentclass[epj]{svjour}

\usepackage{graphicx} \usepackage{psfrag} \usepackage{hyperref}
\usepackage{fancyhdr} \usepackage{amssymb}
\def\makeheadbox{{
 }}
\def\mytitle{My title} 
\def\myauthors{My name}  
\def\mytype{My type of session}
\def\mysession{My session}

\def\mytitle{Inflation and Unification} 
\def\myauthors{Qaisar Shaf\mbox{}i and V. Nefer \c{S}eno\u{g}uz}    
\def\mytype{\mytitle}
\def\mysession{\myauthors}

\begin{document}
\title{Inflation and Unification}
\author{Qaisar Shaf\mbox{}i \and V. Nefer \c{S}eno\u{g}uz\thanks{\emph{Present address:} 
 Department of Physics and Astronomy, University of Kansas, Lawrence, KS 66045, USA}
}                     
%
%
\institute{Bartol Research Institute, Department of Physics and Astronomy, University of Delaware, 
Newark, DE 19716, USA
}
\date{Received: date / Revised version: date}
\date{}
\abstract{Two distinct classes of realistic inflationary models consistent with
present observations are reviewed.
The first example relies on the Coleman-Weinberg potential and is
readily realized within the framework of spontaneously broken global
symmetries (for instance, global $U(1)_{B-L}$). Depending on the parameters
either new or large field inflation is possible.
The second example exploits supersymmetry which makes implementation of
inflation within local gauge theories much more accessible. An example
based on spontaneously broken local $U(1)_{B-L}$ is discussed.
Leptogenesis is naturally realized in both cases.
} 
\maketitle
\section{From new to large field inflation}
An inflationary scenario \cite{Guth:1980zm,Linde:1981mu} 
may be termed successful if it satisfies the following criteria:

1) The total number of $e$-folds N during inflation is large enough to resolve the
horizon and flatness problems. Thus, $N\gtrsim$50--60, but
it can be somewhat smaller for low scale inflation.

2) The predictions are consistent with observations of the microwave
background and large scale structure formation. In particular,
the predictions for $n_s$, $r$ and $\alpha$ should be consistent with the
most recent WMAP results \cite{Spergel:2006hy} (see also \cite{Alabidi:2006qa} for
a brief survey of models).

3) Satisfactory resolution of the monopole problem in grand unified
theories (GUTs) is achieved.

4) Explanation of the origin of the observed baryon asymmetry is provided.

In this section\footnote{Based on Ref. \cite{Shafi:2006cs}.} we review a class of inflation models which appeared in the 
early eighties in the framework of non-supersymmetric GUTs and employed a GUT singlet scalar field
$\phi$ \cite{Shafi:1983bd,Pi:1984pv,Lazarides:1984pq}.
These (Shafi-Vilenkin) models satisfy, as we will see, the above criteria and are based on a
Coleman-Weinberg (CW) potential \cite{Coleman:1973jx}
\begin{equation}
                  V(\phi)= V_0 + A \phi^4 \left[\ln\left(\frac{\phi^2}{M_*^2}\right) + C\right]
\end{equation}
where, following \cite{Shafi:1983bd} the renormalization mass  $M_* = 10^{ 18}$ GeV and
$V_0 ^{1/4}$ will specify the vacuum energy. The value of $C$ is fixed to cancel the cosmological
constant at the minimum. It is convenient to choose a physically
equivalent parametrization for $V(\phi)$ \cite{Albrecht:1984qt,Linde:2005ht}, namely
\begin{equation} \label{pot}
                  V(\phi)= A \phi^4 \left[\ln\left( \frac{\phi}{M}\right) -\frac{1}{4}\right] + \frac{A M^4}{4}\,,
\end{equation}
where $M$ denotes the $\phi$ VEV at the minimum. Note that $V(\phi=M)=0$,
and the vacuum energy density at the origin is given by
                   $V_0 = A M^4 /4$. 
For our discussion here one reasonable choice is to assume
that the global $U(1)_{B-L}$ symmetry of the standard model
is spontaneously broken by the VEV of $\phi$ (see later when we briefly
discuss leptogenesis).

\begin{table*}[t] 
{\centering
\caption{The inflationary parameters for the Shafi-Vilenkin model with the potential in Eq. (\ref{pot}) 
($m_P=1$)} \label{tablo}
\resizebox{!}{2.6cm}{
\begin{tabular}{r@{\hspace{.5cm}}r@{\hspace{.5cm}}r@{\hspace{.5cm}}r@{\hspace{.5cm}}r@{\hspace{.5cm}}r@{\hspace{.5cm}}r@{\hspace{.5cm}}r@{\hspace{.5cm}}r}
\hline \hline
 $V^{1/4}_0$(GeV) & $A(10^{-14})$ & M & $\phi_e$ & $\phi_0$ & $V(\phi_0)^{1/4}$(GeV) & $n_s$ & $\alpha(-10^{-3})$ & $r$   \\
\hline
$10^{13}$ & 1.0 & 0.018 &  0.010 &  $3.0\times10^{-6}$  &$\approx V_0^{1/4}$ &0.938 & 1.4 & $9\times10^{-15}$  \\
\hline
$5\times10^{13}$ & 1.2 & 0.088 &  0.050  & $7.5\times10^{-5}$ &$\approx V_0^{1/4}$ &0.940 & 1.3 & $5\times10^{-12}$   \\
\hline
$10^{14}$ & 1.3 & 0.17 & 0.10  &  $3.0\times10^{-4}$ &$\approx V_0^{1/4}$ &0.940 & 1.2 & $9\times10^{-11}$ \\
\hline
$5\times10^{14}$ & 1.9 & 0.79 & 0.51   & $7.5\times10^{-3}$ & $\approx V_0^{1/4}$&0.941 & 1.2 & $5\times10^{-8}$ \\
\hline
$10^{15}$ & 2.3 & 1.5 & 1.1  & 0.030 &$\approx V_0^{1/4}$ &0.941 & 1.2 & $9\times10^{-7}$   \\
\hline
$5\times10^{15}$ & 4.8 & 6.2 & 5.1  &  0.71 & $\approx V_0^{1/4}$&0.942 & 1.0 & $5\times10^{-4}$ \\
\hline
$10^{16}$ & 5.2 & 12 & 10  &  3.2 & $9.9\times10^{15}$ & 0.952 & 1.0 & $8\times10^{-3}$ \\
\hline
$2\times10^{16}$ & 1.1 & 36 &35   & 23 &$1.7\times10^{16}$ &0.966 & 0.6 & $0.07$ \\
\hline
$3\times10^{16}$ & .17 & 86 &  85  & 72&$1.9\times10^{16}$ &0.967 & 0.6 & $0.11$ \\ 
\hline
$10^{17}$ & .001 & 1035 &  1034  & 1020&$2.0\times10^{16}$ &0.966 & 0.6 & $0.14$ \\
\hline \hline
\end{tabular} }\label{cwtt}
\par} \centering 
\end{table*}

The potential above is typical for the new inflation scenario \cite{Linde:1981mu},
where inflation takes place near the maximum. 
However, as we discuss below, 
depending on the value of $V_0$, the inflaton 
can have small or large values compared to the Planck scale during observable inflation.
In the latter case observable inflation takes place near the minimum and the
model mimics chaotic inflation \cite{Linde:1983gd}.

The original new inflation models attempted to explain the initial value of the inflaton through high-temperature
corrections to the potential. This mechanism does not work unless the inflaton is somewhat small
compared to the Planck scale at the Planck epoch \cite{Linde:2005ht}. However, the initial value of the 
inflaton could also be suppressed by a pre-inflationary phase. Here we will simply assume that the 
initial value of the inflaton is sufficiently small to allow enough $e$-folds.

The slow-roll parameters may be defined as \cite{Liddle:1992wi}
\begin{equation}
\epsilon =
 \frac{1}{2}\left(\frac{V'}{V}\right)^2 \,,\quad
\eta=\left(\frac{V''}{V}\right) \,,\quad
\xi^2=\left(\frac{V'\, V'''}{V^2}\right) \,.
\end{equation}
(Here and below we use units $m_P=1$, where $m_P\simeq2.4\times10^{18}$ GeV
is the reduced Planck mass, although sometimes we will
write $m_P$ explicitly.
The primes denote derivatives with respect to the inflaton $\phi$.) 
The slow-roll approximation is valid if the slow-roll conditions
$\epsilon \ll 1$ and $\eta \ll 1$ hold.
In this case the spectral index
$n_\mathrm{s}$, the tensor to scalar ratio
$r$ and the running of the spectral index
$\alpha\equiv\mathrm{d} n_\mathrm{s}/\mathrm{d} \ln k$ are given by
\begin{eqnarray}
n_\mathrm{s} \!&\simeq&\! 1 - 6 \epsilon + 2 \eta \label{ns}\\
r \!&\simeq&\! 16 \epsilon \\
\alpha \!\!&\simeq&\!\!
16 \epsilon \eta - 24 \epsilon^2 - 2 \xi^2.
\end{eqnarray}

The number of $e$-folds after the comoving scale $l_0=2\pi/k_0$ has crossed the horizon is
given by
\begin{equation} \label{efold0}
N_0=\frac{1}{2}\int^{\phi_0}_{\phi_e}\frac{H(\phi)\rm{d}\phi}{H'(\phi)} \end{equation}
where $\phi_0$ is the value of the field when the scale corresponding to $k_0$
exits the horizon and $\phi_e$ is the value of the field at the end of inflation.
This value is given by the condition $2(H'(\phi)/H(\phi))^2=1$, which can be calculated
from the Hamilton-Jacobi equation \cite{Salopek:1990jq}
\begin{equation}
[H'(\phi)]^2-\frac{3}{2}H^2(\phi)=-\frac{1}{2}V(\phi)\,.
\end{equation}
The amplitude of the
curvature perturbation $\mathcal{P^{{\rm 1/2}}_R}$ is given by
\begin{equation} \label{perturb}
\mathcal{P^{{\rm 1/2}}_R}=\frac{1}{2\sqrt{3}\pi m^3_P}\frac{V^{3/2}}{|V'|}\,.
\end{equation}
To calculate the magnitude of $A$ and the inflationary parameters, we use these standard
equations. We also include the first order corrections in the slow roll expansion
for $\mathcal{P^{{\rm 1/2}}_R}$ and the spectral index $n_s$ \cite{Stewart:1993bc}.\footnote{The 
fractional error in $\mathcal{P^{{\rm 1/2}}_R}$ from the slow
roll approximation is of order $\epsilon$ and $\eta$ (assuming these parameters remain $\ll1$).
This leads to an error in $n_s$ of order $\xi^2$, which is $\sim10^{-3}$ in the present model. 
Comparing to the WMAP errors, this precision seems quite adequate.
However, in anticipation of the Planck mission, it may be desirable
to consider improvements.}
The WMAP value for $\mathcal{P^{{\rm 1/2}}_R}$ is $4.86\times10^{-5}$ for $k_0 = 0.002$ Mpc$^{-1}$.
$N_0$ corresponding to the same scale is $\simeq53+(2/3)\ln(V(\phi_0) ^{1/4}/10^{15}\ \rm{GeV})+(1/3)\ln(T_r/10^{9}\ \rm{GeV})$.
(The expression for $N_0$ assumes a standard thermal history \cite{Dodelson:2003vq}. See \cite{Lyth:1998xn} for reviews.)
We assume reheating is efficient enough such that the reheating temperature $T_r=m_{\phi}$, where the mass of the inflaton
$m_{\phi}=2\sqrt{A}M$. In practice, we expect $T_r$ to be somewhat below $m_{\phi}$ \cite{Shafi:1983bd}.

\begin{figure}[t] 
\psfrag{1-n}{\footnotesize{$1-n_s$}}
\psfrag{r}{\footnotesize{$r$}}
\includegraphics[angle=0, width=8cm]{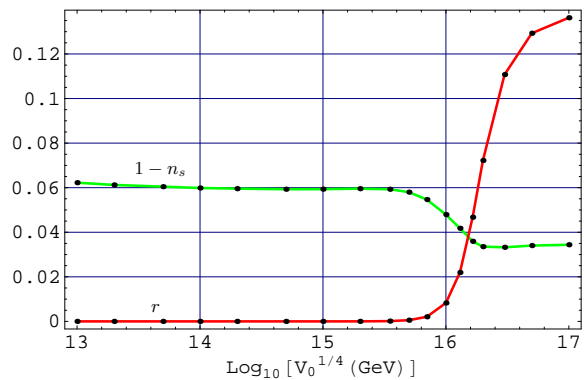} 
\caption{$1-n_s$ and $r$ vs. $\log[V^{1/4}_0{\,\rm(GeV)}]$ for the potential in Eq. (\ref{pot}).} \label{figgg}
\end{figure}

In Table \ref{cwtt}  and Fig. \ref{figgg} we display the predictions for $n_s$, $\alpha$ and $r$, with
the vacuum energy scale $V_0 ^{1/4}$ varying from $10^{13}$ GeV to  $10^{17}$ GeV.
The parameters have a slight dependence on the reheating temperature, as can be seen from
the expression for $N_0$. As an example, if we assume instant reheating ($T_r\simeq V(\phi_0)^{1/4}$), 
$n_s$ would increase to 0.941 and 0.943 for $V_0^{1/4} =10^{13}$ GeV and $V_0^{1/4} =10^{15}$ GeV respectively.

For $V_0^{1/4}\lesssim10^{16}$ GeV, the inflaton field remains smaller than the Planck scale, 
and the inflationary parameters are
similar to those for new inflation models with $V=V_0(1-(\phi/\mu)^4)$: $n_s\simeq1-(3/N_0)$,
$\alpha\simeq (n_s-1)/N_0$.
As the vacuum energy is lowered, $N_0$ becomes smaller and
$n_s$ deviates further from unity. However, $n_s$ remains within $2\sigma$ of the WMAP
best fit value (for negligible $r$) $0.951^{0.015}_{-0.019}$ \cite{Spergel:2006hy} 
even for $V_0^{1/4}$ as low as $10^5$ GeV. Inflation 
with CW potential at low scales is discussed in Ref. \cite{Knox:1992iy}.

\begin{figure}[t] 
\includegraphics[angle=0, width=8cm]{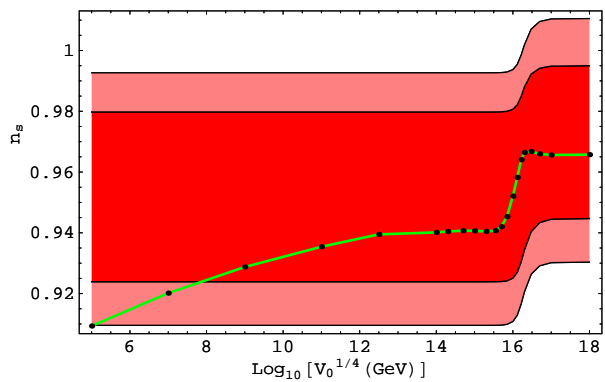} 
\caption{$n_s$ vs. $\log[V^{1/4}_0{\,\rm(GeV)}]$ for the potential in Eq. (\ref{pot}), shown together with 
the WMAP contours (68\% and 95\% confidence levels) \cite{Spergel:2006hy}.} \label{fig:223}
\end{figure}

For $V_0^{1/4}\gtrsim10^{16}$ GeV, the inflaton takes values larger than the Planck scale
during observable inflation. Observable inflation then occurs closer to the minimum 
where the potential is effectively $V=(1/2) m_{\phi}^2 \Delta\phi^2$,
$\Delta\phi=M-\phi$ denoting the deviation of the field from the minimum.
This well-known monomial model \cite{Linde:1983gd} predicts $m_{\phi}\simeq2\times10^{13}$ GeV and
$\Delta\phi_0\simeq2\sqrt{N_0}$, corresponding to $V(\phi_0)\simeq(2\times10^{16}$ GeV$)^4$. 
For the $\phi^2$ potential to be a good approximation, $V_0$ must be greater than 
this value. Then the inflationary parameters no longer depend on $V_0$ and
approach the predictions for the $\phi^2$ potential.

\begin{figure}[t] 
\includegraphics[angle=0, width=8cm]{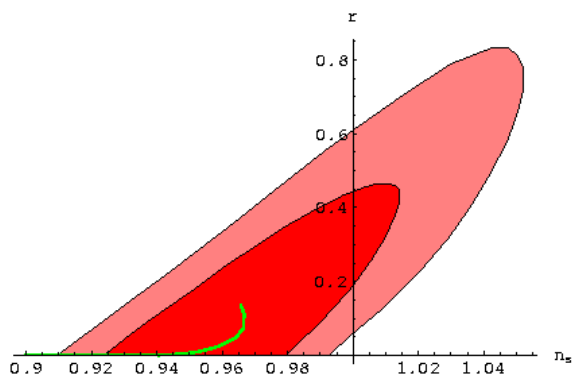} 
\caption{$r$ vs. $n_s$ for the potential in Eq. (\ref{pot}), shown together with 
the WMAP contours (68\% and 95\% confidence levels) \cite{Spergel:2006hy}.} \label{fig:222}
\end{figure}

\begin{figure}[t] 
\includegraphics[angle=0, width=8cm]{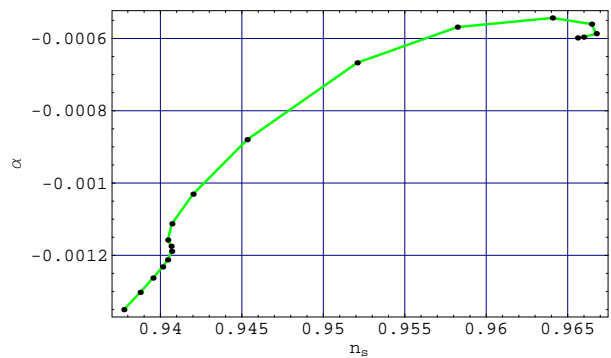} 
\caption{$\alpha$ vs. $n_s$ for the potential in Eq. (\ref{pot}).} \label{fig:2222}
\end{figure}

The spectral index $n_s$ and tensor to scalar ratio $r$ are displayed in Figs. \ref{fig:223}, \ref{fig:222}.
The values are in very good agreement with the recent WMAP results \cite{Spergel:2006hy}. 
The running of the spectral index is negligible, as in most inflation models (Fig. \ref{fig:2222}).

Note that the WMAP analysis suggests a running spectral index, with 
$|\alpha|\lesssim10^{-3}$ disfavored at
the $2\sigma$ level \cite{Spergel:2003cb,Spergel:2006hy}.
On the other hand, an analysis including the constraints
from the Sloan Digital Sky Survey (SDSS) finds no evidence for running
\cite{Seljak:2004xh}.
 Clearly, more data is necessary to resolve this
important issue. Modifications of the models discussed here, generally involving two stages of inflation, 
have been proposed in Refs. \cite{Kawasaki:2003zv,Senoguz:2004ky} and elsewhere to generate a much more significant
variation of $n_s$ with $k$.

In the context of non-supersymmetric GUTs, $V_0^{1/4}$ is related to the unification
scale, and is typically a factor of 3--4 smaller than the superheavy gauge boson masses
due to the loop factor in the CW potential. 
The unification scale for non-supersymmetric GUTs is typically $10^{14}$--$10^{15}$
GeV, although it is possible to have higher scales, for instance associating
inflation with $SO(10)$ breaking via $SU(5)$.

The reader may worry about proton decay with gauge boson masses of order
$10^{14}$--$10^{15}$ GeV.
In the $SU(5)$ model \cite{Georgi:1974sy}, in particular, a two-loop
renormalization group analysis of the standard model gauge couplings yields
masses for the superheavy gauge bosons of order $1\times10^{14}$--$5\times10^{14}$ GeV
\cite{Dorsner:2005ii}. This is consistent with the  SuperK proton lifetime limits \cite{Eidelman:2004wy},
provided one assumes strong flavor suppression of the relevant dimension six
gauge mediated proton decay coefficients. 
If no suppression is assumed the gauge boson masses should have masses
close to $10^{15}$ GeV or higher \cite{Ilia}.

For the Shafi-Vilenkin model
in $SU(5)$, the tree level scalar potential
contains the term $(1/2)\lambda\phi^2{\rm Tr}\Phi^2$ with $\Phi$ being the Higgs adjoint,
and $A\simeq1.5\times10^{-2}\lambda^2$ \cite{Shafi:1983bd,Linde:2005ht}. Inflation requires 
$A\sim10^{-14}$, corresponding to $\lambda\sim10^{-6}$.

This model has been extended to $SO(10)$
in Ref. \cite{Lazarides:1984pq}. The breaking of $SO(10)$ to the standard model proceeds, for
example, via the subgroup $G_{422} = SU(4)_c \times SU(2)_L \times SU(2)_R$ \cite{Pati:1974yy}. A renormalization
group analysis shows that the symmetry breaking scale for $SO(10)$ is of order $10^{15}$
GeV, while $G_{422}$ breaks at an intermediate scale $M_I\sim 10^{12}$ GeV
\cite{Rajpoot:1980xy}. (This is
intriguingly close to the scale needed to resolve the strong CP problem and
produce cold dark matter axions.)
The predictions for $n_s$, $\alpha$ and $r$ are essentially identical
to the SU(5) case. There is one amusing consequence though which may be worth
mentioning here. The monopoles associated with the breaking of $SO(10)$ to $G_{422}$
are inflated away. However, the breaking of $G_{422}$ to
the SM gauge symmetry yields doubly charged monopoles \cite{Lazarides:1980cc}, whose mass is
of order $10^{13}$ GeV. These may be present in our galaxy  at a flux level of
$10^{-16}\ {\rm cm}^{-2}$ s$^{-1}$ sr$^{-1}$ \cite{Lazarides:1984pq}.

As stated earlier, before an inflationary model can be deemed successful, it must contain
a mechanism for generating the observed baryon asymmetry in the universe. In
the $SU(5)$ case the color Higgs triplets produced by
inflaton decay can generate the baryon asymmetry, provided  the Higgs sector of
the model has the required amount of CP violation \cite{Shafi:1983bd}.

The discovery of neutrino oscillations requires that we introduce SU(5)
singlet right handed neutrinos, presumably three of them, to implement
the seesaw mechanism and generate the desired masses for the light
neutrinos. In this case it is natural to generate the observed baryon
asymmetry via leptogenesis \cite{Fukugita:1986hr} 
by introducing the couplings
                   $N_i N_j \phi^2 / m_P$,
where $N_i$ (i=1,2,3) denote the right handed neutrinos, and the renormalizable
coupling to $\phi$ is absent because of the assumed discrete
symmetry. By suitably adjusting the Yukawa coefficients one can arrange
that the $\phi$ field decays into the right handed neutrinos.
Note that the presence of the above Yukawa couplings then allows one
to make the color triplets heavier, of order $10^{14}$ GeV, thereby avoiding
any potential conflict with proton decay. In the $SO(10)$ model,
leptogenesis is almost automatic \cite{Lazarides:1984pq}.

\section{$U(1)_{B-L}$: Neutrino Physics and Inflation}
Physics beyond the Standard Model (SM) is required by the following experimental
observations:

\begin{itemize}
\item Neutrino Oscillations: $\Delta m_{\rm SM}^2 \lesssim 10^{-10}\ {\rm eV}^2\ll$ (mass difference)$^2$
                          needed to understand atmospheric and solar neutrino
                          observations;
\item CMB Anisotropy ($\delta T/T$): requires inflation which cannot be realized in the
                             SM;
\item Non-Baryonic Dark Matter ($\Omega_{\rm CDM} = 0.25$): SM has no plausible candidate;
\item Baryon Asymmetry ($n_b/s\sim10^{-10}$): Not possible to achieve in the SM.
\end{itemize}

Recall that at the renormalizable level the SM possesses a global $U(1)_{B-L}$
symmetry. If the symmetry is gauged, anomaly cancellation requires the
existence of three right handed neutrinos. An important question therefore
is the symmetry breaking scale of $U(1)_{B-L}$. Note that this scale is not fixed
by the evolution of the three SM gauge couplings. Remarkably, we will be able
to determine the $M_{B-L}$ by implementing inflation. With $M_{B-L}$ well below the
Planck scale the seesaw mechanism enables us to realize light neutrino masses
in the desired range. Furthermore, it will turn out that leptogenesis is a
natural outcome after inflation is over.

The introduction of a gauge $U(1)_{B-L}$ symmetry broken at a scale well below the
Planck scale exacerbates the well known gauge hierarchy problem. There are at
least four potential hierarchy problems one could consider:
\begin{itemize}
\item $M_W \ll M_P$;
\item $M_{B-L} \ll M_P$ (required by neutrino oscillations);
\item $m_{\chi} \ll M_P$ (where $m_{\chi}$ denotes the inflaton mass);
\item $f_a \sim 10^{10}-10^{12}\ {\rm GeV}\ (\ll M_P)$, where $f_a$ denotes the axion decay constant.
\end{itemize}
Supersymmetry (SUSY) can certainly help here, especially if the SUSY breaking
scale in the observable sector is of order TeV. Thus, it seems that a good
starting point, instead of SM$\,\times U(1)_{B-L}$, could be MSSM$\,\times$\\$U(1)_{B-L}$.
The $Z_2$ `matter' parity associated with the MSSM has two important consequences. It
eliminates rapid (dimension four) proton decay, and it delivers a respectable
cold dark matter candidate in the form of LSP. However, Planck scale suppressed
dimension five proton decay is still present and one simple solution is to
embed $Z_2$ in a $U(1)_R$ symmetry. It turns out that the R symmetry also plays an
essential role in realizing a compelling inflationary scenario and in the
resolution of the MSSM $\mu$ problem. Finally it seems natural to extend the above
discussion to larger groups, especially to $SO(10)$ and its various subgroups.

\subsection{Supersymmetric Hybrid Inflation Models}
\label{chap2}
In this section\footnote{Based on \cite{Senoguz:2004vu}.} we review 
a class of supersymmetric hybrid inflation models \cite{Lazarides:2001zd}
where inflation can be linked to the breaking of $U(1)_{B-L}$.
We compute the allowed range of the
dimensionless coupling in the superpotential and the dependence of the
spectral index on this coupling, in the presence of canonical supergravity (SUGRA)
corrections.

The simplest
supersymmetric hybrid inflation model \cite{Dvali:1994ms} is realized by the
renormalizable superpotential \cite{Copeland:1994vg}
\begin{equation} \label{super} W_1=\kappa S(\Phi\overline{\Phi}-M^{2})
\end{equation}
\noindent where $\Phi(\overline{\Phi})$ denote a conjugate pair of superfields
transforming as nontrivial representations of some gauge group $G$, $S$ is a
gauge singlet superfield, and $\kappa$ $(>0)$ is a dimensionless coupling.  A
suitable $U(1)$ R-symmetry, under which $W_1$ and $S$ transform the same way,
ensures the uniqueness of this superpotential at the renormalizable level \cite{Dvali:1994ms}.
In the absence of supersymmetry breaking, the potential
energy minimum corresponds to non-zero vacuum expectation values (VEVs) $(=M)$ in the
scalar right handed neutrino components
$\big|\langle\nu^c_H\rangle\big|=\big|\langle\overline{\nu}^c_H\rangle\big|$ for
$\Phi$ and $\overline{\Phi}$, while the VEV of $S$ is
zero.  (We use the same notation for superfields and their scalar components.)
Thus, $G$ is broken to some subgroup $H$ which, in many interesting models,
coincides with the MSSM gauge group.

In order to realize inflation, the scalar fields $\Phi$,
$\overline{\Phi}$, $S$ must be displayed from their present minima.  For
$|S|>M$, the $\Phi$, $\overline{\Phi}$ VEVs both vanish so that the gauge
symmetry is restored, and the tree level potential energy density
$\kappa^{2}M^{4}$ dominates the universe, as in the originally
proposed hybrid inflation scenario \cite{Linde:1991km,Copeland:1994vg}.
With supersymmetry thus broken, there are radiative corrections from the
$\Phi-\overline{\Phi}$ supermultiplets that provide logarithmic corrections
to the potential which drives inflation.

In one loop approximation the inflationary effective potential is given
by \cite{Dvali:1994ms}
{\setlength\arraycolsep{2pt}
\begin{eqnarray} \label{loop} V_{\mathrm{LOOP}}=\kappa^{2}M^{4}\bigg[1
+\frac{\kappa^{2}\mathcal{N}}{32\pi^{2}} \Big(
2\ln\frac{\kappa^{2}|S|^{2}}{\Lambda^{2}}+\nonumber\\(z+1)^{2}\ln(1+z^{-1}) 
+(z-1)^{2}\ln(1-z^{-1})\Big) \bigg]\,, \end{eqnarray}}
\noindent where $z\equiv x^{2}\equiv|S|^{2}/M^{2}$, $\mathcal{N}$ is the dimensionality of
the $\Phi$, $\overline{\Phi}$ representations, and $\Lambda$ is
a renormalization mass scale.

The scalar spectral index $n_s$ is given by Eq. (\ref{ns}),
where primes denote derivatives with respect to the normalized
real scalar field $\sigma\equiv\sqrt{2}|S|$. For relevant values of the
parameters ($\kappa\ll1$), the slow roll conditions ($\epsilon$, $\eta\ll1$) are violated only
`infinitesimally'
close to the critical point at $x=1$ ($|S|=M$) \cite{Lazarides:2001zd}.
So inflation continues practically until this point is
reached, where it abruptly ends.

The number of $e$-folds after the comoving scale $l_0$ has crossed the horizon is
given by Eq. (\ref{efold0}), which in the slow roll approximation can also be written as
\begin{equation} \label{efold1}
N_0=\frac{1}{m^2_P}\int^{\sigma_0}_{\sigma_e}\frac{V\rm{d}\sigma}{V'}\,. \end{equation}
Using Eqs. (\ref{loop}, \ref{efold1}),
we obtain
\begin{equation} \label{m_kap}
\kappa\approx\frac{2\sqrt{2}\pi}{\sqrt{\mathcal{N}N_{0}}}\,y_{0}\,\frac{M}{m_{P}}\,.
\end{equation}
(The subscript 0 implies that the values correspond to $k_0\equiv0.002$ Mpc$^{-1}$.)
$N_0\approx55$ is the number of $e$-folds and
\begin{equation} \label{yq} y_{0}^{2}=\int_{1}^{x^{2}_{0}}\frac{\textrm{d} z}{z
f(z)}\quad ,y_0\ge 0\,,  \end{equation}
with
\begin{equation} \label{feza}
f(z)=\left(z+1\right)\ln\left(1+z^{-1}\right)+\left(z-1\right)\ln\left(1-z^{-1}\right)\,.
\end{equation}
Using Eqs. (\ref{loop}, \ref{m_kap}, \ref{perturb}),
$\mathcal{P^{{\rm 1/2}}_R}$ is found to be \cite{Dvali:1994ms,Lazarides:2001zd,Lazarides:1997dv}
\begin{equation} \label{quad}
\mathcal{P^{{\rm 1/2}}_R}\approx2\left(\frac{N_0}{3\mathcal{N}}\right)^{1/2}
\left(\frac{M}{m_{P}}\right)^{2}x_{0}^{-1}y_{0}^{-1}f (x^{2}_{0})^{-1}\,.
\end{equation}

Up to now, we ignored supergravity (SUGRA) corrections to the potential.
More often than not, SUGRA corrections tend to derail an otherwise succesful
inflationary scenario by giving rise to scalar (mass)$^2$ terms of order
$H^2$, where $H$ denotes the Hubble constant. Remarkably, it turns out that for
a canonical SUGRA potential (with minimal K\"ahler potential
$|S|^2+|\Phi|^2+|\overline{\Phi}|^2$), the problematic (mass)$^2$ term cancels
out for the superpotential $W_1$ in Eq. (\ref{super}) \cite{Copeland:1994vg}.
This property also persists when non-renormalizable terms that are permitted by
the $U(1)_R$ symmetry are included in the superpotential.

In general,
$K$ can be expanded as
\begin{equation} \label{kappas}
K=|S|^2+|\Phi|^2+|\overline{\Phi}|^2+\kappa_S\frac{|S|^4}{4m^2_P}+\ldots\,,
\end{equation}
and only the
$|S|^4$ term in $K$ generates a (mass)$^2$ for $S$, which would spoil
inflation for $\kappa_S\sim1$ \cite{Panagiotakopoulos:1997ej}.\footnote{We 
should note that, since the superpotential is linear in the inflaton, 
the presence of other fields with Planck scale VEVs
would also induce an inflaton mass of order $H$. Some ways to suppress the inflaton mass
are discussed in \cite{Panagiotakopoulos:2004tf}.}

The scalar potential is given by
\begin{equation} \label{SUGRA}
V={\rm e}^K\left[\left(\frac{\partial^2 K}{\partial z_i\partial z^*_j}\right)^{-1}D_{z_i}W D_{z^*_j}W^*-3|W|^2\right]+V_D\,,
\end{equation}
with
\begin{equation} \label{SUGRA2}
D_{z_i}W=\frac{\partial W}{\partial z_i}+\frac{\partial K}{\partial z_i}W\,,
\end{equation}
where the sum extends over all fields $z_i$, and $K=\sum_i |z_i|^2$
is the minimal K\"ahler potential.  The D-term $V_D$ vanishes in the D-flat direction
$|\overline{\Phi}|=|\Phi|$. From Eq. (\ref{SUGRA}), with a minimal K\"ahler potential one contribution
to the inflationary potential is given by \cite{Copeland:1994vg,Panagiotakopoulos:1997qd,Linde:1997sj,Kawasaki:2003zv}
\begin{equation}
V_{\rm SUGRA}=\kappa^{2}M^{4}\left[\frac{|S|^{4}}{2}+\ldots\right]\,.
\end{equation}
There are additional contributions to the potential arising from the soft
SUSY breaking terms. In $N=1$ SUGRA these include the universal scalar masses
equal to $m_{3/2}$ ($\sim$ TeV), the gravitino mass.
However, their effect on the
inflationary scenario is negligible, as discussed below.
The more important term is the $A$ term
 $(2-A)m_{3/2}\kappa M^2 S(+\rm{h.c.})$. For convenience, we write this as
$a\,m_{3/2} \kappa M^2 |S|$, where $a\equiv2|2-A|\cos(\arg S+\arg(2-A))$.
The effective potential is approximately given by
Eq. (\ref{loop}) plus the leading SUGRA correction $\kappa^2 M^4 |S|^4/2$ and
the $A$ term:
{\setlength\arraycolsep{2pt}
\begin{eqnarray} \label{potential} V_{1}&=&\kappa^{2}M^{4}\bigg[1
+\frac{\kappa^{2}\mathcal{N}}{32\pi^{2}} \Big(
2\ln\frac{\kappa^{2}|S|^{2}}{\Lambda^{2}}\nonumber\\&+&(z+1)^{2}\ln(1+z^{-1})
+(z-1)^{2}\ln(1-z^{-1})\Big)\nonumber\\&+&\frac{|S|^4}{2} \bigg]+ a\,m_{3/2} \kappa M^2|S|\,. \end{eqnarray}}
We perform our numerical calculations using this potential, taking $|a\,m_{3/2}|$=1 TeV.
It is, however, instructive to discuss small and large $\kappa$ limits of Eq. (\ref{potential}).
For $\kappa\gg10^{-3}$, $1\gg\sigma\gg \sqrt{2}M$, and Eq. (\ref{potential}) becomes
\begin{equation} \label{v1}
V_1\simeq\kappa^{2}M^{4}\left[1+\frac{\kappa^{2}\mathcal{N}}{32\pi^{2}}2\ln\frac{\kappa^{2}\sigma^{2}}{2\Lambda^{2}}+
\frac{\sigma^{4}}{8}\right] \end{equation}
to a good approximation. Comparing the derivatives of the radiative and SUGRA corrections
one sees that the radiative term dominates for
$\sigma^2\lesssim\kappa\sqrt{\mathcal{N}}/2\pi$. From $3H\dot{\sigma}=-V'$,
$\sigma^2_0\simeq\kappa^2 \mathcal{N}N_0/4\pi^2$ for the one-loop effective
potential, so that SUGRA effects are negligible only for
$\kappa\ll2\pi/\sqrt{\mathcal{N}}N_{0}\simeq 0.1/\sqrt{\mathcal{N}}$. (For
$\mathcal{N}=1$, this essentially agrees with \cite{Linde:1997sj}.)

$\mathcal{P^{{\rm 1/2}}_R}$ is found from Eq. (\ref{v1}) to be
\begin{equation} \mathcal{P^{{\rm 1/2}}_R}\simeq\frac{1}{\sqrt{3}\pi}\frac{\kappa\,M^2}{\sigma^3_0}\,.  \end{equation}
\noindent In the absence of the SUGRA correction, the gauge symmetry breaking
scale $M$ is given by Eq. (\ref{quad}). For $\kappa\gg10^{-3}$, $x_0\gg1$ and
$x_{0}\,y_{0}\,f(x^{2}_{0})\to1^{-}$.  $\mathcal{P^{{\rm 1/2}}_R}$ in this case 
turns out to be proportional to $(M/m_{P})^2$\cite{Dvali:1994ms,Lazarides:2001zd}.  
Using the WMAP best fit $\mathcal{P^{{\rm 1/2}}_R}\simeq4.7\times10^{-5}$ \cite{Spergel:2003cb}, 
$M$ approaches the value $\mathcal{N}^{1/4}\cdot6\times10^{15}$ GeV. The presence of
the SUGRA term leads to larger values of $\sigma_0$ and hence larger values of
$M$ for $\kappa\gtrsim0.06/\sqrt{\mathcal{N}}$.

For $\kappa\ll10^{-3}$, $|S_0|\simeq M$ where $S_0$ is the value
of the field at $k_0$, i.e. $z\simeq1$.  (Note that due to the extreme flatness of the
potential the last 55 or so $e$-folds occur with $|S|$ close to $M$.)
From Eqs. (\ref{perturb}, \ref{potential}), as $z\to1$
\begin{equation} \label{spert}
\mathcal{P^{{\rm 1/2}}_R}=\frac{2\sqrt{2}\pi}{\sqrt{3}}\frac{\kappa^2 M^4}{\ln(2)\kappa^3 M \mathcal{N}+8\pi^2 \kappa M^5
+4\pi^2 a\,m_{3/2}}\,.
\end{equation}
The denominator of Eq. (\ref{spert}) contains the radiative, SUGRA and the $A$
terms respectively.  Comparing them, we see that the radiative term can be
ignored for $\kappa\lesssim10^{-4}$. There is also a soft mass term $m^2_{3/2} |S|^2$ in the potential, 
corresponding to an additional term \\ $8\pi^2 m^2_{3/2} /\kappa M$ in the denominator.
We have omitted this term, since it is insignificant
for $\kappa\gtrsim10^{-5}$.

For a positive $A$ term ($a>0$), the maximum value of $\mathcal{P^{{\rm 1/2}}_R}$
as a function of $M$ is found to be
\begin{equation} \label{rmax}
\mathcal{P^{{\rm 1/2}}_R}_{\rm{max}}=\frac{1}{2^{7/10}\, 5^{3/2}\, 3\pi}\left(\frac{\kappa^6}{a\,m_{3/2}}\right)^{1/5}\,.
\end{equation}
Setting $\mathcal{P^{{\rm
1/2}}_R}\simeq4.7\times10^{-5}$, we find a
lower bound on $\kappa$ ($\simeq10^{-5}$). For larger values of
$\kappa$, there are two separate solutions of $M$ for a given
$\kappa$. The solution with larger $M$ is not valid if the symmetry breaking pattern produces cosmic strings. 
For example, strings are produced when $\Phi,\,\overline{\Phi}$ break $U(1)_R\times U(1)_{B-L}$ to $U(1)_Y\times Z_2$ 
matter parity, but not when $\Phi,\,\overline{\Phi}$ are $SU(2)_R\times U(1)_{B-L}$ doublets.
(These two examples correspond to $\mathcal{N}=1$ and $\mathcal{N}=2$ respectively.)
For $a<0$, there are again two solutions, but for the solution with
a lower value of $M$, the slope changes sign as the inflaton rolls
for $\kappa\lesssim10^{-4}$ and the inflaton gets trapped in a false
vacuum.

Note that the $A$ term depends on $\arg S$, so it
should be checked whether $\arg S$ changes significantly during inflation.
Numerically, we find that it does not, except for a range of $\kappa$ around
$10^{-4}$ \cite{Senoguz:2004vu}. For this range,
if the initial value of the $S$ field is greater than $M$ by at least a factor of two or so,
the $A$ term and the slope become negative even if they were initially positive, before inflation can
suitably end. However, larger values of the $A$ term, or the mass term coming from a non-minimal K\"ahler potential
(or from a hidden sector VEV) would drive the value of $M$ in that region up, allowing the slope to stay positive
(see Ref. \cite{Jeannerot:2005mc} for the effect of varying the $A$ term and the mass).

\begin{figure}[t]
\psfrag{k}{\footnotesize{$\kappa$}}
\includegraphics[height=.2\textheight]{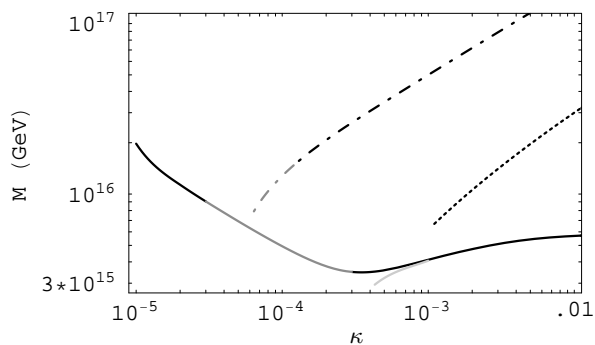}
\caption{  The value of the symmetry breaking scale $M$ vs.
$\kappa$, for SUSY hybrid inflation with $\mathcal{N}=1$ (solid),
and for shifted hybrid inflation
(dot-dashed for $M_S=m_P$, dotted for $M_S=5\times10^{17}$ GeV).
Light grey portions of the curves are for $a<0$, where only the
segments that do not overlap with the solutions for $a>0$ are shown.
The grey segments denote the range of $\kappa$ for
which the change in $\arg S$ is significant.} \label{fig:21}
\end{figure}

The dependence of $M$ on $\kappa$ is shown in Fig. \ref{fig:21}. 
Note that with inflation linked to the breaking of MSSM$\,\times$\\$ U(1)_{B-L}$, $M$ 
corresponds to the $U(1)_{B-L}$ breaking scale, which
is not fixed by the evolution of the three SM gauge couplings.
The amplitude of the curvature perturbation (or, equivalently, $\delta T/T$)
determines this scale to be close to the SUSY GUT scale, suggesting that
$U(1)_{B-L}$ could be embedded in $SO(10)$ or its subgroups. For example,
$M$ can be determined in flipped $SU(5)$
                from the renormalization group evolution of the $SU(3)$ and
                $SU(2)$ gauge couplings. The values are remarkably consistent
                with the ones fixed from $\delta T/T$ considerations \cite{Kyae:2005nv}.

Here, some remarks concerning the allowed range of $\kappa$ is in order.
As discussed above, a lower bound on $\kappa$ is obtained from the inflationary dynamics
and the amplitude of the curvature perturbation.
An upper bound on $\kappa$ is obtained from the value of the spectral index, which
we discuss next. The gravitino constraint provides a more stringent upper bound ($\kappa\lesssim10^{-2}$), as discussed
in the next section. If cosmic strings form, 
the range of $\kappa$ is also restricted by the limits on the cosmic string contribution 
to $\mathcal{P^{{\rm 1/2}}_R}$, however most of the range may still be allowed \cite{Jeannerot:2005mc,Battye:2006pk}.

\begin{figure}[t]
\psfrag{k}{\footnotesize{$\kappa$}}
\includegraphics[height=.2\textheight]{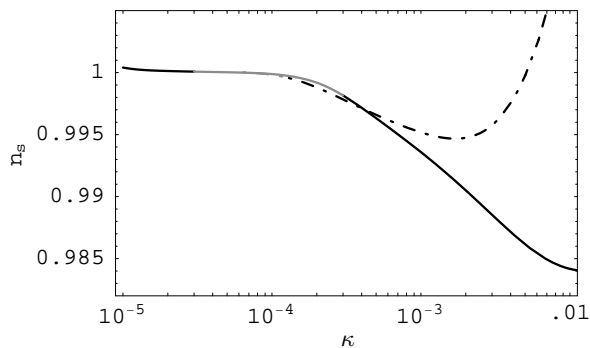}
\caption{  The spectral index $n_s$ vs. $\kappa$, for SUSY hybrid
inflation with $\mathcal{N}=1$ (solid), and for shifted hybrid inflation with $M_S=m_P$
(dot-dashed). The grey segments denote the range of $\kappa$ for
which the change in $\arg S$ is significant.} \label{fig:22}
\end{figure}

\begin{figure}[t]
\psfrag{k}{\footnotesize{$\kappa$}}
\includegraphics[height=.2\textheight]{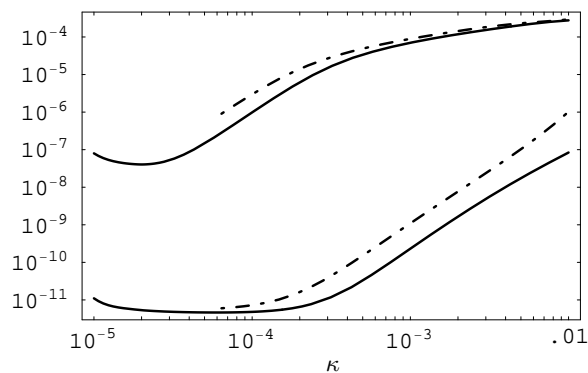}
\caption{$\alpha$ vs. $\kappa$ (top) and tensor to scalar ratio $r$ vs. $\kappa$ (bottom), 
for SUSY hybrid inflation with $\mathcal{N}=1$ (solid), and for shifted hybrid inflation with $M_S=m_P$
(dot-dashed).} \label{fig:rdn}
\end{figure}

In the absence of SUGRA corrections, the scalar spectral index $n_s$
for $\kappa\gg10^{-3}$ is given by \cite{Dvali:1994ms}
\begin{equation} n_{s}\simeq1+2\eta\simeq 1-\frac{1}{N_{0}}\simeq0.98\,,
\end{equation}
while it approaches unity for small $\kappa$.
When the SUGRA correction is taken into account,
one finds that the spectral index $n_s$ exceeds unity for
$\kappa\simeq2\pi/\sqrt{3\mathcal{N}}N_0\simeq0.06/\sqrt{\mathcal{N}}$ \cite{Senoguz:2003zw}.  The
dependence of $n_s$ on $\kappa$ is displayed in Fig. \ref{fig:22}.
$\alpha$ is small and the tensor to scalar ratio $r$ is negligible,
as shown in Fig. \ref{fig:rdn}.

\begin{figure}[htb]
\hspace{-.5cm}\includegraphics[height=.3\textheight]{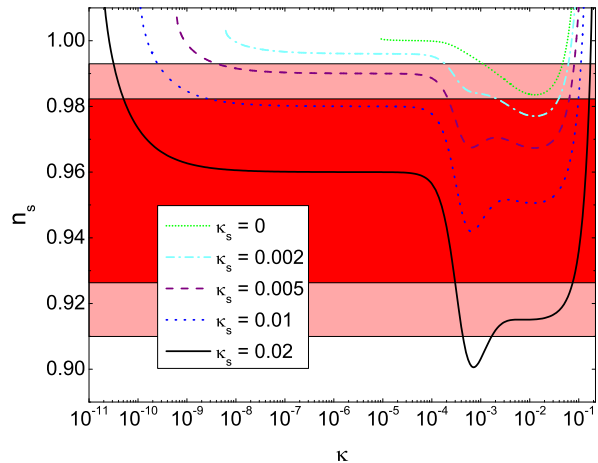}
\vspace{-.7cm}
\caption{The spectral index $n_s$ vs. $\kappa$, for SUSY hybrid
inflation with $\mathcal{N}=1$ and different values of $\kappa_S$ \cite{ur Rehman:2006hu}.} \label{fig:242}
\end{figure}

For negligible $r$, the WMAP three year central value for the spectral index is 
$n_{s}\approx0.95$, and SUSY hybrid inflation with a minimal K\"ahler potential
is disfavoured at a $2\sigma$ level \cite{Spergel:2006hy}.
It was recently shown that the spectral index for SUSY hybrid inflation can be
in better agreement with the WMAP3 results in the presence of a small negative mass term in the
potential. This can result from a non-minimal K\"ahler potential, in particular
from the term proportional to the dimensionless coupling $\kappa_S$ referred to in Eq. (\ref{kappas})
\cite{Boubekeur:2005zm,ur Rehman:2006hu}.
The spectral index $n_s$ for different values of $\kappa_S$ is displayed in 
Fig. \ref{fig:242}.\footnote{Note that the cosmic string contribution is not
included in Fig.  \ref{fig:242}. Inclusion of cosmic strings change the WMAP
contours, allowing larger values of $n_s$ depending on the string tension.
In particular, SUSY hybrid inflation with a minimal K\"ahler potential ($\kappa_S=0$)
can then provide a good fit to WMAP data
for $\mathcal{N}=1$ and $\kappa\simeq10^{-2}$ \cite{Battye:2006pk,Bevis:2007gh}.
On the other hand, cosmic strings would be absent for other symmetry breaking
patterns such as the $\mathcal{N}=2$ example mentioned above. Also, cosmic strings
are inflated away in the shifted hybrid inflation model discussed below.}

The inflationary scenario based on the superpotential $W_1$ in
Eq. (\ref{super}) has the characteristic feature that the end of
inflation essentially coincides with the gauge symmetry breaking.  Thus,
modifications should be made to $W_1$ if the breaking of $G$ to $H$ leads to
the appearance of topological defects such as monopoles or domain
walls. For instance, the breaking of $G_{PS}\equiv SU(4)_c\times SU(2)_L\times
SU(2)_R$ \cite{Pati:1974yy} to the MSSM by fields belonging to
$\Phi(\overline{4},1,2)$, $\overline{\Phi}(4,1,2)$ produces magnetic monopoles
that carry two quanta of Dirac magnetic charge \cite{Lazarides:1980cc}. As
shown in \cite{Jeannerot:2000sv}, one simple resolution of the topological defects problem
is achieved by supplementing $W_1$ with a non-renormalizable term:
\begin{equation} \label{super2} W_2=\kappa
S(\overline{\Phi}\Phi-v^{2})-\frac{S(\overline{\Phi}\Phi)^{2}}{M^{2}_{S}}\,,
\end{equation}
\noindent where $v$ is comparable to the SUSY GUT scale
$M_{\rm GUT}\simeq2\times10^{16}$ GeV and $M_{S}$ is an effective
cutoff scale. The dimensionless coefficient of the non-renormalizable term is
absorbed in $M_S$.  The presence of the non-renormalizable term enables an
inflationary trajectory along which the gauge symmetry is broken.  Thus, in
this `shifted' hybrid inflation model the topological defects are inflated away.

The inflationary potential is similar to Eq. (\ref{potential}):
{\setlength\arraycolsep{2pt}
\begin{eqnarray} \label{potential2} V_{2}&=&
\kappa^{2}m^{4}\bigg[1+\frac{\kappa^{2}}{16\pi^{2}} \Big(
2\ln\frac{\kappa^{2}|S|^{2}}{\Lambda^{2}} \\\nonumber&+&(z+1)^{2}\ln(1+z^{-1}) +(z-1)^{2}\ln(1-z^{-1})\Big) 
\\\nonumber &+&\frac{|S|^4}{2}\bigg] +a\,m_{3/2}\kappa v^2|S|
+\kappa^3 m^4 M^2_S |S|^2{}\,.  \end{eqnarray}}
Here $m^{2}=v^{2}(1/4\xi-1)$ where $\xi=v^{2}/\kappa M^{2}_{S}$, $z\equiv x^{2}\equiv\sigma^{2}/m^{2}$,
and $2-A$ is replaced by $2-A+A/2\xi$ in the expression for $a$. Note 
that the potential also contains a mass term even for minimal K\"ahler potential, 
due to the nonvanishing VEVs of $\Phi,\,\overline{\Phi}$.

The VEV
$M=\big|\langle\nu^c_H\rangle\big|=\big|\langle\overline{\nu}^c_H\rangle\big|$ at
the SUSY minimum is given by \cite{Jeannerot:2000sv}
\begin{equation}
\left(\frac{M}{v}\right)^{2}=\frac{1}{2\xi}\left(1-\sqrt{1-4\xi}\right)\,,
\end{equation}
\noindent and is $\sim10^{16}-10^{17}$ GeV depending on $\kappa$ and $M_S$.
The system follows the inflationary trajectory for
$1/7.2<\xi<1/4$, which is satisfied for
$\kappa\gtrsim10^{-5}$ if the effective cutoff scale $M_S=m_P$.
For lower values of $M_S$, the inflationary trajectory is followed only for
higher values of $\kappa$, and $M$ is lower for a given $\kappa$ (Fig. \ref{fig:21}).
The spectral index is displayed in Fig. \ref{fig:22}.

\subsection{Leptogenesis In Supersymmetric Hybrid Inflation Models}
\label{chap3}
An important constraint on SUSY hybrid inflation models
arises from considering the reheating temperature $T_{r}$
after inflation, taking into account the gravitino problem which requires that
$T_{r}\lesssim10^6$--$10^{11}$ GeV \cite{Khlopov:1984pf}. This
constraint on $T_r$ depends on the SUSY breaking mechanism and the
gravitino mass $m_{3/2}$. For gravity mediated SUSY breaking models
with unstable gravitinos of mass $m_{3/2}\simeq0.1$--1 TeV,
$T_r\lesssim10^6$--$10^9$ GeV \cite{Kawasaki:1995af}, while
$T_r\lesssim10^{10}$ GeV for stable gravitinos \cite{Bolz:2000fu}.  In gauge
mediated models the reheating temperature is generally more severely constrained, although
$T_r\sim10^9$--$10^{10}$ GeV is possible for $m_{3/2}\simeq5$--100 GeV
\cite{Gherghetta:1998tq}. Finally, the anomaly mediated symmetry breaking (AMSB)
scenario may allow gravitino masses much heavier than a TeV, thus
accommodating a reheating temperature as high as $10^{11}$ GeV \cite{Gherghetta:1999sw}.

After the end of inflation in the models discussed in section~\ref{chap2}, the
fields fall toward the SUSY vacuum and perform damped oscillations about it.
The vevs of $\overline{\Phi}$, $\Phi$ along their right handed neutrino
components $\overline{\nu}^c_H$, $\nu^c_H$  break the gauge symmetry.  The
oscillating system, which we collectively denote as $\chi$, consists of the two
complex scalar fields $(\delta\overline{\nu}^c_H+\delta\nu^c_H)/\sqrt{2}$
(where $\delta\overline{\nu}^c_H$, $\delta\nu^c_H$ are the deviations of
$\overline{\nu}^c_H$, $\nu^c_H$ from $M$) and $S$, with equal mass $m_{\chi}$.

We assume here that the inflaton $\chi$ decays predominantly into right handed neutrino
superfields $N_i$, via the superpotential coupling $(1/m_P)\gamma_{ij}
\overline{\phi}\,\overline{\phi}N_i N_j$ or $\gamma_{ij}\overline{\phi} N_i N_j$,
where $i,j$ are family indices 
(see below for a different scenario connected to the resolution of the MSSM $\mu$
problem). Their subsequent out of equilibrium decay to lepton and Higgs
superfields generates lepton asymmetry, which is then partially converted into
the observed baryon asymmetry by sphaleron effects
\cite{Fukugita:1986hr}. 

The right handed neutrinos, as shown below, can be heavy compared to
the reheating temperature $T_r$.
Unlike thermal leptogenesis,
there is then no supression factor in the lepton asymmetry, 
since the washout is proportional to the Boltzmann factor $e^{-z}$ (where $z=M_1/T_r$) 
and can be neglected for $z\gtrsim10$ \cite{Buchmuller:2003gz}.\footnote{Lepton number violating 2-body scatterings
mediated by right handed neutrinos are also out of equilibrium \cite{Giudice:1999fb}.}
Without this assumption, generating sufficient lepton asymmetry
would require a reheat temperature $\gtrsim2\times10^9$ GeV
\cite{Giudice:2003jh}.

\begin{figure}[t] 
\psfrag{m}{\scriptsize{$m_{\chi}$ (GeV)}}
\psfrag{k}{\footnotesize{$\kappa$}}
\includegraphics[height=.2\textheight]{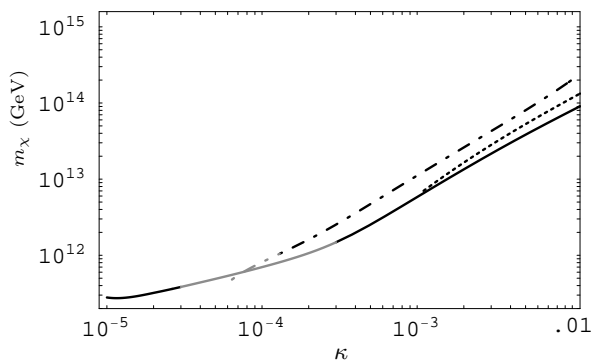} 
\caption{\sf The inflaton mass $m_{\chi}$ vs. $\kappa$,  
for SUSY hybrid inflation with $\mathcal{N}=1$ (solid), 
and for shifted hybrid inflation 
(dot-dashed for $M_S=m_P$, dotted for $M_S=5\times10^{17}$ GeV). 
The grey segments denote the range of $\kappa$ for which the change in $\arg S$ is significant.} \label{minf}
\end{figure}

\begin{figure}[t] 
\psfrag{k}{\footnotesize{$\kappa$}}
\includegraphics[height=.2\textheight]{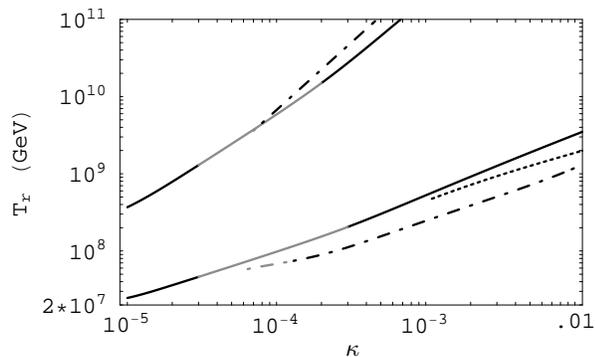} 
\caption{\sf The lower bound on the reheating temperature $T_r$ vs. $\kappa$,  
for SUSY hybrid inflation with $\mathcal{N}=1$ (solid) and for shifted hybrid inflation 
(dot-dashed for $M_S=m_P$, dotted for $M_S=5\times10^{17}$ GeV). The segments in the top left part of the figure correspond to the bounds in the presence of a $\lambda S h^2$ coupling. The grey segments denote the range of $\kappa$ for which the change in $\arg S$ is significant.} \label{ktr}
\end{figure}

GUTs typically relate the Dirac neutrino masses to that of the quarks or charged leptons.
It is therefore reasonable to assume the Dirac masses are hierarchical. The
low-energy neutrino data indicates that the right handed neutrinos
will then also be hierarchical in general.  A reasonable mass pattern is  $M_1<M_2\ll M_3$, which can result from either
the dimensionless couplings $\gamma_{ij}$ or additional symmetries (see e.g. \cite{Senoguz:2004ky}).
The dominant contribution to the lepton asymmetry is
still from the decays with $N_3$ in the loop, as long as the first two family
right handed neutrinos are not quasi degenerate.
The lepton asymmetry is then given by \cite{Asaka:1999yd}
\begin{equation} \label{nls}
\frac{n_L}{s}\lesssim3\times10^{-10}\frac{T_r}{m_{\chi}}\left(\frac{M_i}{10^6\rm{\
GeV}}\right)\left(\frac{m_{\nu3}}{0.05\rm{\ eV}}\right)\,, \end{equation}
where $M_i$ denotes the mass of the heaviest right handed neutrino the inflaton
can decay into. 
The decay rate $\Gamma_{\chi}=(1/8\pi)(M^2_i/M^2)m_{\chi}$
\cite{Lazarides:1997dv}, and the reheating temperature $T_r$ is given by
\begin{equation} \label{reheat} T_r=\left(\frac{45}{2\pi^2
g_*}\right)^{1/4}(\Gamma_\chi\,m_P)^{1/2}\simeq
0.063\frac{(m_P\,m_{\chi})^{1/2}}{M}M_i \,.\end{equation}
From the experimental value of the
baryon to photon ratio $\eta_B\simeq6.1\times10^{-10}$ \cite{Spergel:2003cb},
the required lepton asymmetry is found to be $n_L/s\simeq2.5\times10^{-10}$
\cite{Khlebnikov:1988sr}.
Using this value, along with Eqs. (\ref{nls}, \ref{reheat}), we can express $T_r$ in terms of
the symmetry breaking scale $M$ and the inflaton mass $m_{\chi}$:
\begin{eqnarray} \label{trmin}
T_r&\gtrsim&\left(\frac{10^{16}{\rm\
GeV}}{M}\right)^{1/2}\left(\frac{m_{\chi}}{10^{11}{\rm\ GeV}}\right)^{3/4}\nonumber\\
&\times&\left(\frac{0.05\rm{\ eV}}{m_{\nu3}}\right)^{1/2}
1.6\times10^{7}{\rm\ GeV}
\,.  \end{eqnarray}
Here $m_{\chi}$ is given by $\sqrt{2}\kappa M$ and $\sqrt{2}\kappa M \sqrt{1-4\xi}$ 
respectively for hybrid and shifted hybrid inflation.
The value of $m_{\chi}$ is shown in Fig. \ref{minf}.
We show the lower bound on $T_r$ calculated using this equation (taking $m_{\nu3}=0.05$ eV) 
in Fig. \ref{ktr}. 

Eq. (\ref{reheat}) also yields the result that the heaviest right handed
neutrino the inflaton can decay into is about 400 (6) times heavier than
$T_r$, for hybrid inflation with $\kappa=10^{-5}$ ($10^{-2}$). For shifted hybrid inflation, this ratio
does not depend on $\kappa$ as strongly and
is $\sim10^2$. This is consistent with ignoring washout
effects as long as the lightest right handed neutrino mass $M_1$ is also $\gg T_r$.\footnote{If
$M_1<T_r$, part of the asymmetry created by decays of the next-to-lightest right handed neutrino
will be washed out. However, the asymmetry that survives the washout can still
be sufficient to account for the observed baryon asymmetry \cite{Senoguz:2007hu}.}

Both the gravitino constraint and the constraint $M_1\gg T_r$ favor smaller
values of $\kappa$ for hybrid inflation, with
$T_r\gtrsim2\times10^7$ GeV for $\kappa\sim10^{-5}$. 
Similarly, the gravitino constraint favors $\kappa$ values as small as the inflationary trajectory
allows for shifted hybrid inflation, and $T_r\gtrsim5\times10^7$ GeV for $M_S=m_P$.

So far we have not addressed the $\mu$ problem and the relationship to $T_r$ in the present context.
The MSSM $\mu$ problem can naturally be resolved in SUSY hybrid inflation models
in the presence of the term $\lambda Sh^2$ in the superpotential, where
$h$ contains the two Higgs doublets \cite{Dvali:1998uq}. (The `bare' term $h^2$ is not allowed
by the $U(1)$ R-symmetry.) After inflation
the VEV of $S$ generates a $\mu$ term with $\mu=\lambda\langle S\rangle=-m_{3/2}\lambda/\kappa$,
where $\lambda>\kappa$ is required for the proper vacuum.
The inflaton in this case predominantly decays into higgses (and higgsinos) with $\Gamma_h=(1/16\pi)
\lambda^2 m_{\chi}$. As a consequence the presence of this term
significantly increases the reheating temperature $T_r$. Following Ref. \cite{Lazarides:1998qx},
we calculate $T_r$ for the best case scenario
$\lambda=\kappa$. We find  a lower bound on $T_r$ of $4\times10^{8}$ GeV in hybrid inflation,
see Fig. \ref{ktr}. $T_r\gtrsim4\times10^9$ GeV for shifted hybrid inflation with
$M_S=m_P$.
An alternative resolution of the $\mu$ problem in SUSY hybrid inflation involves
a Peccei-Quinn (PQ) symmetry $U(1)_{\rm PQ}$ \cite{Lazarides:1998iq,Jeannerot:2000sv}.

The lower bounds on $T_r$ are summarized in Table \ref{ttable}. 
There is some tension between the gravitino constraint and the
reheating temperature required to generate sufficient lepton asymmetry,
particularly for gravity mediated SUSY breaking models, and if hadronic decays
of gravitinos are not suppressed.
However, we should note that 
having quasi degenerate neutrinos would increase the lepton asymmetry per neutrino
decay $\epsilon$ \cite{Flanz:1996fb} and thus allow lower values of $T_r$ corresponding to lighter right
handed neutrinos. Provided that the neutrino mass splittings are comparable to
their decay widths, $\epsilon$ can be as large as $1/2$ \cite{Pilaftsis:1997jf}.
The lepton asymmetry in this case is of order $T_r/m_{\chi}$ where $m_{\chi}\sim10^{11}$
GeV for $\kappa\sim10^{-5}$, and sufficient lepton asymmetry
can be generated with $T_r$ close to the electroweak scale.

\begin{table}[t]
\caption{Lower bounds on the reheating temperature (GeV)}
\begin{tabular}{l@{\hspace{.5cm}}r@{\hspace{.5cm}}r}
\hline \hline
 & without $\lambda S h^2$ & with $\lambda S h^2$ \\
\hline
SUSY hybrid inflation & $2\times10^7$ & $4\times10^8$ \\
\hline
Shifted hybrid inflation & $5\times10^7$ & $4\times10^9$ \\
\hline \hline
\end{tabular}
\label{ttable}
\end{table}

Finally, it is worth noting that new inflation models have also been 
considered in the framework of supersymmetric GUTs, taking account of
supergravity corrections. In Ref. \cite{Senoguz:2004ky}, for instance, it is
shown that the spectral index $n_s$ is less than 0.98.  Furthermore, reheating temperatures as low
as $10^4$--$10^6$ GeV can be realized to satisfy the gravitino constraint.

\subsection*{Acknowledgements}
This work is partially supported by the US Department of Energy under contract number  DE-FG02-91ER40626.


\begin{thebibliography}{99}

\bibitem{Guth:1980zm}
  A.~H.~Guth,
  Phys.\ Rev.\  D {\bf 23}, 347 (1981).

\bibitem{Linde:1981mu}
  A.~D.~Linde,
  Phys.\ Lett.\  B {\bf 108}, 389 (1982);
  A.~Albrecht and P.~J.~Steinhardt,
  Phys.\ Rev.\ Lett.\  {\bf 48}, 1220 (1982).

\bibitem{Spergel:2006hy}
  D.~N.~Spergel {\it et al.}  [WMAP Collaboration],
  Astrophys.\ J.\ Suppl.\  {\bf 170}, 377 (2007)
  [arXiv:astro-ph/0603449].

\bibitem{Alabidi:2006qa}
  L.~Alabidi and D.~H.~Lyth,
  JCAP {\bf 0608}, 013 (2006)
  [arXiv:astro-ph/0603539].

\bibitem{Shafi:2006cs}
  Q.~Shafi and V.~N.~Senoguz,
  Phys.\ Rev.\  D {\bf 73}, 127301 (2006)
  [arXiv:astro-ph/0603830].

\bibitem{Shafi:1983bd}
  Q.~Shafi and A.~Vilenkin,
  Phys.\ Rev.\ Lett.\  {\bf 52}, 691 (1984).

\bibitem{Pi:1984pv}
  S.~Y.~Pi,
  Phys.\ Rev.\ Lett.\  {\bf 52}, 1725 (1984);
  Q.~Shafi and A.~Vilenkin,
  Phys.\ Rev.\  D {\bf 29}, 1870 (1984).

\bibitem{Lazarides:1984pq}
  G.~Lazarides and Q.~Shafi,
  Phys.\ Lett.\  B {\bf 148}, 35 (1984).

\bibitem{Coleman:1973jx}
  S.~R.~Coleman and E.~Weinberg,
  Phys.\ Rev.\  D {\bf 7}, 1888 (1973).



\bibitem{Albrecht:1984qt}
  A.~Albrecht and R.~H.~Brandenberger,
  Phys.\ Rev.\  D {\bf 31}, 1225 (1985);
  A.~Albrecht, R.~H.~Brandenberger and R.~Matzner,
  Phys.\ Rev.\  D {\bf 32}, 1280 (1985).

\bibitem{Linde:2005ht}
  A.~D.~Linde,
  arXiv:hep-th/0503203.

\bibitem{Linde:1983gd}
  A.~D.~Linde,
  Phys.\ Lett.\ {\bf B129}, 177 (1983).

\bibitem{Liddle:1992wi}
  A.~R.~Liddle and D.~H.~Lyth,
  Phys.\ Lett.\  B {\bf 291}, 391 (1992)
  [arXiv:astro-ph/9208007].

\bibitem{Salopek:1990jq}
  D.~S.~Salopek and J.~R.~Bond,
  Phys.\ Rev.\  D {\bf 42}, 3936 (1990).

\bibitem{Stewart:1993bc}
  E.~D.~Stewart and D.~H.~Lyth,
  Phys.\ Lett.\  B {\bf 302}, 171 (1993)
  [arXiv:gr-qc/9302019].

\bibitem{Dodelson:2003vq}
  S.~Dodelson and L.~Hui,
  Phys.\ Rev.\ Lett.\  {\bf 91}, 131301 (2003)
  [arXiv:astro-ph/0305113];
  A.~R.~Liddle and S.~M.~Leach,
  Phys.\ Rev.\ {\bf D68}, 103503 (2003)
  [arXiv:astro-ph/0305263].

\bibitem{Lyth:1998xn}
  D.~H.~Lyth and A.~Riotto,
  Phys.\ Rept.\  {\bf 314}, 1 (1999)
  [arXiv:hep-ph/9807278];
  A.~R.~Liddle and D.~H.~Lyth,
{\it  Cambridge, UK: Univ. Pr. (2000) 400 p}


\bibitem{Knox:1992iy}
  L.~Knox and M.~S.~Turner,
  Phys.\ Rev.\ Lett.\  {\bf 70}, 371 (1993)
  [arXiv:astro-ph/9209006].


\bibitem{Spergel:2003cb}
  D.~N.~Spergel {\it et al.}  [WMAP Collaboration],
  Astrophys.\ J.\ Suppl.\  {\bf 148}, 175 (2003)
  [arXiv:astro-ph/0302209];
  H.~V.~Peiris {\it et al.}  [WMAP Collaboration],
  Astrophys.\ J.\ Suppl.\  {\bf 148}, 213 (2003)
  [arXiv:astro-ph/0302225].

\bibitem{Seljak:2004xh}
  U.~Seljak {\it et al.},
  Phys.\ Rev.\ D {\bf 71}, 103515 (2005)
  [arXiv:astro-ph/0407372];
  U.~Seljak, A.~Slosar and P.~McDonald,
  JCAP {\bf 0610}, 014 (2006)
  [arXiv:astro-ph/0604335].

\bibitem{Kawasaki:2003zv}
  M.~Kawasaki, M.~Yamaguchi and J.~Yokoyama,
  Phys.\ Rev.\  D {\bf 68}, 023508 (2003)
  [arXiv:hep-ph/0304161].

\bibitem{Senoguz:2004ky}
  V.~N.~Senoguz and Q.~Shafi,
  Phys.\ Lett.\  B {\bf 596}, 8 (2004)
  [arXiv:hep-ph/0403294].

\bibitem{Georgi:1974sy}
  H.~Georgi and S.~L.~Glashow,
  Phys.\ Rev.\ Lett.\  {\bf 32}, 438 (1974).

\bibitem{Dorsner:2005ii}
  I.~Dorsner, P.~F.~Perez and R.~Gonzalez Felipe,
  Nucl.\ Phys.\  B {\bf 747}, 312 (2006)
  [arXiv:hep-ph/0512068].

\bibitem{Eidelman:2004wy}
  S.~Eidelman {\it et al.}  [Particle Data Group],
  Phys.\ Lett.\ {\bf B592}, 1 (2004).


\bibitem{Ilia}
I. Dorsner, private communication.


\bibitem{Pati:1974yy}
  J.~C.~Pati and A.~Salam,
  Phys.\ Rev.\  D {\bf 10}, 275 (1974)
  [Erratum-ibid.\  D {\bf 11}, 703 (1975)].

\bibitem{Rajpoot:1980xy}
  S.~Rajpoot,
  Phys.\ Rev.\  {\bf D22}, 2244 (1980);
  G.~Lazarides, Q.~Shafi and C.~Wetterich,
  Nucl.\ Phys.\  {\bf B181}, 287 (1981);
  D.~G.~Lee, R.~N.~Mohapatra, M.~K.~Parida and M.~Rani,
  Phys.\ Rev.\  D {\bf 51}, 229 (1995)
  [arXiv:hep-ph/9404238];
  F.~Buccella and D.~Falcone,
  Mod.\ Phys.\ Lett.\  A {\bf 18}, 1819 (2003)
  [arXiv:hep-ph/0304143].


\bibitem{Lazarides:1980cc}
  G.~Lazarides, M.~Magg and Q.~Shafi,
  Phys.\ Lett.\  B {\bf 97}, 87 (1980).


\bibitem{Fukugita:1986hr}
  M.~Fukugita and T.~Yanagida,
  Phys.\ Lett.\  B {\bf 174}, 45 (1986).
 For non-thermal
leptogenesis, see 
  G.~Lazarides and Q.~Shafi,
  Phys.\ Lett.\  B {\bf 258}, 305 (1991).


\bibitem{Senoguz:2004vu}
  V.~N.~Senoguz and Q.~Shafi,
  Phys.\ Rev.\  D {\bf 71}, 043514 (2005)
  [arXiv:hep-ph/0412102];
  arXiv:hep-ph/0512170.

\bibitem{Lazarides:2001zd}
For a review and additional references, see  G.~Lazarides,
  Lect.\ Notes Phys.\  {\bf 592}, 351 (2002)
  [arXiv:hep-ph/0111328].

\bibitem{Dvali:1994ms}
  G.~R.~Dvali, Q.~Shafi and R.~K.~Schaefer,
  Phys.\ Rev.\ Lett.\  {\bf 73}, 1886 (1994)
  [arXiv:hep-ph/9406319].

\bibitem{Copeland:1994vg}
  E.~J.~Copeland, A.~R.~Liddle, D.~H.~Lyth, E.~D.~Stewart and D.~Wands,
  Phys.\ Rev.\  D {\bf 49}, 6410 (1994)
  [arXiv:astro-ph/9401011].

\bibitem{Linde:1991km}
  A.~D.~Linde,
  Phys.\ Lett.\  B {\bf 259}, 38 (1991);
  A.~R.~Liddle and D.~H.~Lyth,
  Phys.\ Rept.\  {\bf 231}, 1 (1993)
  [arXiv:astro-ph/9303019].

\bibitem{Lazarides:1997dv}
  G.~Lazarides, R.~K.~Schaefer and Q.~Shafi,
  Phys.\ Rev.\  D {\bf 56}, 1324 (1997)
  [arXiv:hep-ph/9608256].

\bibitem{Panagiotakopoulos:1997ej}
  C.~Panagiotakopoulos,
  Phys.\ Lett.\  B {\bf 402}, 257 (1997)
  [arXiv:hep-ph/9703443].

\bibitem{Panagiotakopoulos:2004tf}
  C.~Panagiotakopoulos,
  Phys.\ Rev.\  D {\bf 71}, 063516 (2005)
 [arXiv:hep-ph/0411143].

\bibitem{Panagiotakopoulos:1997qd}
  C.~Panagiotakopoulos,
  Phys.\ Rev.\  D {\bf 55}, 7335 (1997)
  [arXiv:hep-ph/9702433].

\bibitem{Linde:1997sj}
  A.~D.~Linde and A.~Riotto,
  Phys.\ Rev.\  D {\bf 56}, 1841 (1997)
  [arXiv:hep-ph/9703209].



\bibitem{Jeannerot:2005mc}
  R.~Jeannerot and M.~Postma,
  JHEP {\bf 0505}, 071 (2005)
  [arXiv:hep-ph/0503146].


\bibitem{Kyae:2005nv}
  B.~Kyae and Q.~Shafi,
  Phys.\ Lett.\  B {\bf 635}, 247 (2006)
  [arXiv:hep-ph/0510105].


\bibitem{Battye:2006pk}
  R.~A.~Battye, B.~Garbrecht and A.~Moss,
  JCAP {\bf 0609}, 007 (2006)
  [arXiv:astro-ph/0607339].

\bibitem{Senoguz:2003zw}
  V.~N.~Senoguz and Q.~Shafi,
  Phys.\ Lett.\  B {\bf 567}, 79 (2003)
  [arXiv:hep-ph/0305089].

\bibitem{Boubekeur:2005zm}
  L.~Boubekeur and D.~H.~Lyth,
  JCAP {\bf 0507}, 010 (2005)
  [arXiv:hep-ph/0502047];
  M.~Bastero-Gil, S.~F.~King and Q.~Shafi,
  Phys.\ Lett.\  B {\bf 651}, 345 (2007)
  [arXiv:hep-ph/0604198].

\bibitem{ur Rehman:2006hu}
  M.~ur Rehman, V.~N.~Senoguz and Q.~Shafi,
  Phys.\ Rev.\  D {\bf 75}, 043522 (2007)
  [arXiv:hep-ph/0612023].

\bibitem{Bevis:2007gh}
  N.~Bevis, M.~Hindmarsh, M.~Kunz and J.~Urrestilla,
  arXiv:astro-ph/0702223.

\bibitem{Jeannerot:2000sv}
  R.~Jeannerot, S.~Khalil, G.~Lazarides and Q.~Shafi,
  JHEP {\bf 0010}, 012 (2000)
  [arXiv:hep-ph/0002151].

\bibitem{Khlopov:1984pf}
  M.~Y.~Khlopov and A.~D.~Linde,
  Phys.\ Lett.\  B {\bf 138}, 265 (1984).

\bibitem{Kawasaki:1995af}
  M.~Kawasaki and T.~Moroi,
  Prog.\ Theor.\ Phys.\  {\bf 93}, 879 (1995)
  [arXiv:hep-ph/9403364].

\bibitem{Bolz:2000fu}
  M.~Bolz, A.~Brandenburg and W.~Buchmuller,
  Nucl.\ Phys.\  B {\bf 606}, 518 (2001)
  [arXiv:hep-ph/0012052];
%
  M.~Fujii, M.~Ibe and T.~Yanagida,
  Phys.\ Lett.\  B {\bf 579}, 6 (2004)
  [arXiv:hep-ph/0310142];
%
  L.~Roszkowski, R.~Ruiz de Austri and K.~Y.~Choi,
  JHEP {\bf 0508}, 080 (2005)
  [arXiv:hep-ph/0408227].

\bibitem{Gherghetta:1998tq}
  T.~Gherghetta, G.~F.~Giudice and A.~Riotto,
  Phys.\ Lett.\  B {\bf 446}, 28 (1999)
  [arXiv:hep-ph/9808401].

\bibitem{Gherghetta:1999sw}
  T.~Gherghetta, G.~F.~Giudice and J.~D.~Wells,
  Nucl.\ Phys.\  B {\bf 559}, 27 (1999)
  [arXiv:hep-ph/9904378].

\bibitem{Buchmuller:2003gz}
  W.~Buchmuller, P.~Di Bari and M.~Plumacher,
  Nucl.\ Phys.\  B {\bf 665}, 445 (2003)
  [arXiv:hep-ph/0302092].

\bibitem{Giudice:1999fb}
  G.~F.~Giudice, M.~Peloso, A.~Riotto and I.~Tkachev,
  JHEP {\bf 9908}, 014 (1999)
  [arXiv:hep-ph/9905242].

\bibitem{Giudice:2003jh}
  G.~F.~Giudice, A.~Notari, M.~Raidal, A.~Riotto and A.~Strumia,
  Nucl.\ Phys.\  B {\bf 685}, 89 (2004)
  [arXiv:hep-ph/0310123];
  W.~Buchmuller, P.~Di Bari and M.~Plumacher,
  Annals Phys.\  {\bf 315}, 305 (2005)
  [arXiv:hep-ph/0401240].

\bibitem{Asaka:1999yd}
  T.~Asaka, K.~Hamaguchi, M.~Kawasaki and T.~Yanagida,
  Phys.\ Lett.\  B {\bf 464}, 12 (1999)
  [arXiv:hep-ph/9906366].

\bibitem{Khlebnikov:1988sr}
  S.~Y.~Khlebnikov and M.~E.~Shaposhnikov,
  Nucl.\ Phys.\  B {\bf 308}, 885 (1988).

\bibitem{Senoguz:2007hu}
  V.~N.~Senoguz,
  Phys.\ Rev.\  D {\bf 76}, 013005 (2007)
  [arXiv:0704.3048 [hep-ph]].

\bibitem{Dvali:1998uq}
  G.~R.~Dvali, G.~Lazarides and Q.~Shafi,
  Phys.\ Lett.\  B {\bf 424}, 259 (1998)
  [arXiv:hep-ph/9710314];
%
  S.~F.~King and Q.~Shafi,
  Phys.\ Lett.\  B {\bf 422}, 135 (1998)
  [arXiv:hep-ph/9711288].

\bibitem{Lazarides:1998qx}
  G.~Lazarides and N.~D.~Vlachos,
  Phys.\ Lett.\  B {\bf 441}, 46 (1998)
  [arXiv:hep-ph/9807253].

\bibitem{Lazarides:1998iq}
  G.~Lazarides and Q.~Shafi,
  Phys.\ Rev.\  D {\bf 58}, 071702 (1998)
  [arXiv:hep-ph/9803397].

\bibitem{Flanz:1996fb}
  M.~Flanz, E.~A.~Paschos, U.~Sarkar and J.~Weiss,
  Phys.\ Lett.\  B {\bf 389}, 693 (1996)
  [arXiv:hep-ph/9607310].

\bibitem{Pilaftsis:1997jf}
  A.~Pilaftsis,
  Phys.\ Rev.\  D {\bf 56}, 5431 (1997)
  [arXiv:hep-ph/9707235].



\end{thebibliography}
\end{document}